\documentclass[aps,prb,twocolumn,superscriptaddress,footinbib,floatfix,showkeys,preprint,10pt,longbibliography]{revtex4-2}
\usepackage{graphicx}  % Include figure files
\usepackage{subfigure}
\usepackage{dcolumn} % Align table columns on decimal point
\usepackage{bm}% bold math
\usepackage{amssymb}   % for math
\usepackage{amsfonts}   % for math
\usepackage{comment}
\usepackage[colorlinks,linkcolor=black,anchorcolor=black,citecolor=black,urlcolor=black]{hyperref}
\usepackage{xcolor}
\usepackage{soul}
\usepackage{tabularx,booktabs,ragged2e,multirow}
\newcolumntype{C}{>{\centering\arraybackslash}X}
\usepackage{multirow}
\usepackage{amsmath}
\usepackage{mathrsfs}
\usepackage{mathcomp}
\usepackage{textcomp}
\usepackage{dsfont}
\usepackage{esint}
\usepackage{braket}
\usepackage{textgreek}
\usepackage{lipsum}
\usepackage{marvosym} 
\usepackage{centernot}
\usepackage{cancel}
\usepackage{dashrule}
\begin{document}
%\preprint{APS/123-QED}

%\title{Low-energy signature and magnetic suppression of NH skin effects}
\title{Anisotropic three-dimensional quantum Hall effect and magnetotransport in mesoscopic Weyl semimetals}

\author{Xiao-Xiao Zhang}
\email{xiaoxiao.zhang@riken.jp}
\affiliation{RIKEN Center for Emergent Matter Science (CEMS), Wako, Saitama 351-0198, Japan}

\author{Naoto Nagaosa}
\email{nagaosa@ap.t.u-tokyo.ac.jp}
\affiliation{Department of Applied Physics, University of Tokyo, Tokyo 113-8656, Japan}
\affiliation{RIKEN Center for Emergent Matter Science (CEMS), Wako, Saitama 351-0198, Japan}

%\date{\today}

\newcommand{\br}{{\bm r}}
\newcommand{\bk}{{\bm k}}
\newcommand{\bq}{{\bm q}}
\newcommand{\bp}{{\bm p}}
\newcommand{\bv}{{\bm v}}
\newcommand{\bmm}{{\bm m}}
\newcommand{\bA}{{\bm A}}
\newcommand{\bcA}{{\bm {\mathcal A}}}
\newcommand{\bE}{{\bm E}}
\newcommand{\bB}{{\bm B}}
\newcommand{\cB}{{\mathcal B}}
\newcommand{\bcB}{{\bm {\mathcal B}}}
\newcommand{\bH}{{\bm H}}
\newcommand{\bP}{{\bm P}}
\newcommand{\bd}{{\bm d}}
\newcommand{\bzero}{{\bm 0}}
\newcommand{\bOmega}{{\bm \Omega}}
\newcommand{\bsigma}{{\bm \sigma}}
\newcommand{\bX}{{\bm X}}
\newcommand{\bx}{{\bm x}}
\newcommand{\bS}{{\bm S}}
\newcommand\dd{\mathrm{d}}
\newcommand\ii{\mathrm{i}}
\newcommand\ee{\mathrm{e}}
\newcommand\zz{\mathtt{z}}
\makeatletter
\let\newtitle\@title
\let\newauthor\@author
\def\ExtendSymbol#1#2#3#4#5{\ext@arrow 0099{\arrowfill@#1#2#3}{#4}{#5}}
\newcommand\LongEqual[2][]{\ExtendSymbol{=}{=}{=}{#1}{#2}}
\newcommand\LongArrow[2][]{\ExtendSymbol{-}{-}{\rightarrow}{#1}{#2}}
\newcommand{\cev}[1]{\reflectbox{\ensuremath{\vec{\reflectbox{\ensuremath{#1}}}}}}
\newcommand{\red}[1]{\textcolor{red}{#1}} %for displaying red texts
\newcommand{\blue}[1]{\textcolor{blue}{#1}} %for displaying blue texts
\newcommand{\green}[1]{\textcolor{orange}{#1}} %for displaying blue texts
\newcommand{\mytitle}[1]{\textcolor{orange}{\textit{#1}}}
\newcommand{\mycomment}[1]{} %for commenting out
\newcommand{\note}[1]{ \textbf{\color{blue}#1}}
\newcommand{\warn}[1]{ \textbf{\color{red}#1}}

\makeatother

\begin{abstract}
Weyl semimetals are emerging to become a new class of quantum-material platform for various novel phenomena. Especially, the Weyl orbit made from surface Fermi arcs and bulk relativistic states is expected to play a key role in magnetotransport, leading even to a three-dimensional quantum Hall effect (QHE). It is experimentally and theoretically important although yet unclear whether it bears essentially the same phenomenon as the conventional two-dimensional QHE. 
We discover an unconventional fully three-dimensional anisotropy in the quantum transport under magnetic field. Strong suppression and even disappearance of QHE occur when Hall-bar current is rotated away from being transverse to parallel with respect to the Weyl point alignment, which is attributed to a peculiar absence of conventional bulk-boundary correspondence. Besides, transport along the magnetic field can exhibit a remarkable reversal from negative to positive magnetoresistance. These results establish the uniqueness of this QHE system as a novel three-dimensional quantum matter.

\end{abstract}
%\pacs{71.10.Pm, 71.27.+a, 72.15.Nj, 72.15.Rn}

\maketitle
% \tableofcontents
% \newpage
% \clearpage

%\clearpage
% \let\oldaddcontentsline\addcontentsline% Store \addcontentsline
% \renewcommand{\addcontentsline}[3]{}% Make \addcontentsline a no-op

\section*{Introduction}

\begin{figure}[hbt]
\includegraphics[width=8.6cm]{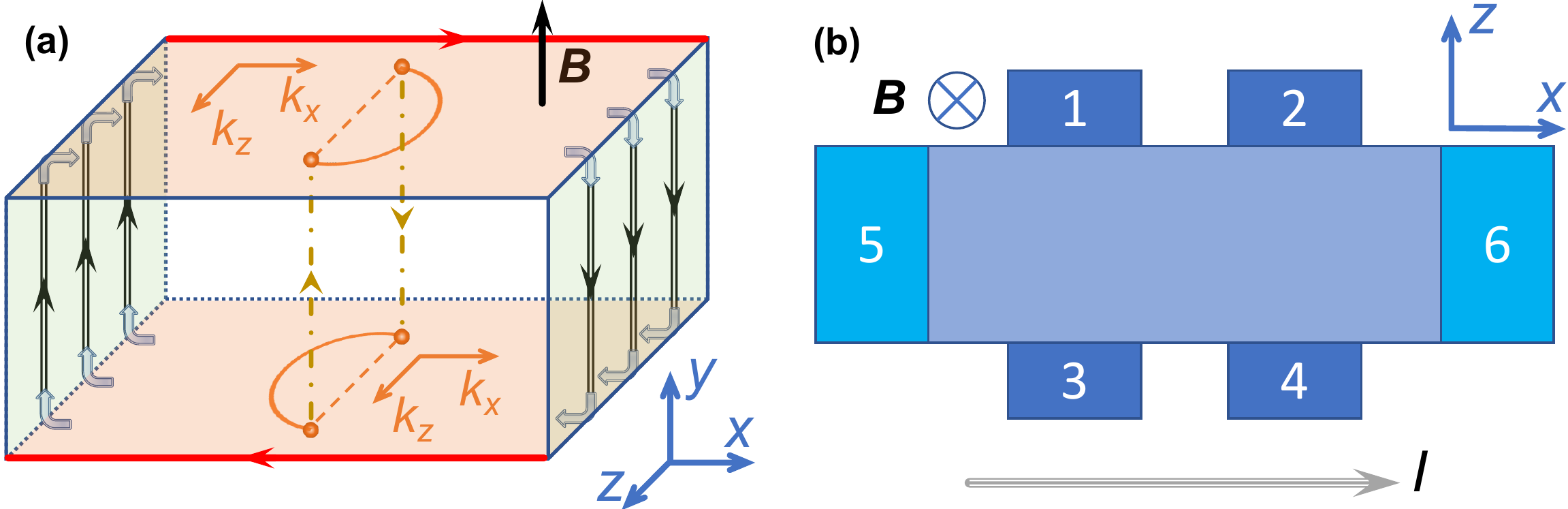}
\caption{(a) The 3D bulk WSM system under magnetic field. The orange top and bottom surfaces at $y=\pm L_y/2$ are always connected by the green side surfaces at $x=\pm L_x/2$ due to the topological arc circulating states. The red arc LL edge states lead to the robust QHE. To illustrate the Weyl orbit, we use a mixed representation: on the top and bottom surfaces we indicate in $\bk$-space the Fermi arcs connecting two WPs along $k_z$; the brown dot-dashed lines along real $y$-axis indicate the bulk chiral LLs. Note that such chiral LLs exist everywhere in $\br$-space. (b) 6-terminal Hall-bar system under magnetic field $\bB=B\hat{y}$. Leads 1,2,3,4 measure voltage while a current flows through leads 5,6. 2-terminal case has leads 5,6 only. The axes exemplify the $zx$-geometry with QHE signal.}\label{Fig:HallBar}
\end{figure}

Continuous interest has been sparked by the prediction and experimental realization of three-dimensional (3D) linear band crossings in Weyl semimetal (WSM) systems as a highly nontrivial extension of the two-dimensional (2D) Dirac physics\cite{Volovik1987,Weyl2007,Weyl2011,AHE2,WeylwithP1,TaAs1,TaAs2,TaAs3,TaAs4}. WSM possesses various intriguing features such as i) topologically protected momentum-space monopole structure, ii) open-arc surface states, iii) bulk chiral Landau level (LL) formation, iv) anomalous Hall effect (AHE), and v) chiral magnetic effect and negative magnetoresistance\cite{Nielson-Ninomiya2,CME1,Burkov2014,seeCMEDirac1,seeCMEDirac2,seeCMEWeyl2,Liang2018,Zhang2019a,Volovik,Burkov2016,ReviewYan2017,Burkov2018,WeylDiracReview,Nagaosa_2020,Lv2021}. 
In this new quantum material, the unique 3D closed semiclassical Weyl orbit under magnetic field has been proposed: it consists of a combination of the top and bottom surface Fermi arcs and the bulk chiral LLs, where the latter leads to a real-space vertical trajectory connecting the projection points of a Weyl point (WP) on the orange surfaces in Fig.~\ref{Fig:HallBar}(a)\cite{Potter2014,Zhang2016b,Zhang2021a}. Apart from its detection from quantum phase oscillations, researchers also envisaged a 3D quantum Hall effect (QHE) based on such orbit in mesoscopic WSMs\cite{Wang2017a}. 
Previous studies focused on the %semiclassical picture 
dependence of magnetic field direction and clarified other possible mechanisms\cite{Li2020a,Chen_2021,Chen_2021a,Chang_2021}; experimental investigation in related materials also found encouraging evidence of the Weyl orbit\cite{Zhang2017,Uchida2017,Schumann2018,Zhang2018,Lin2019a,Nishihaya2021,Zhang2021a}. 

\begin{table*}[hbt]%The best place to locate the table environment is directly after its first reference in text
\centering
\begin{tabular}{c c | c | c }
 %& & \multicolumn{2}{c}{intrinsic}  \\
%\cline{2-5}
%\cmidrule(lr){3-4} %\cmidrule(lr){5-5}
%\cline{3-4} \cline{5-5}
% & & $\chi$ & $\nu$   \\
%\cline{2-4}
%\hline
%\cline{2-5}
%\cmidrule{2-4}  
\multirow{2}{*}{$\bB=B\hat{y}$} & \multicolumn{3}{c}{current direction $i $} \\
\cmidrule{2-4}
& $x$ (QHE)** & \multicolumn{1}{c|}{$y$*} & \multicolumn{1}{c}{$z$} \\
\cmidrule{1-4} 
%\cline{3-6}
major contribution & diagonal edge states from arc LLs & $\textrm{arc channels}+\textrm{bulk chiral LLs}$ & $\textrm{surface hybridized states}$ \\
\cmidrule{1-4} 
2-terminal conductance $G_{i i }/\frac{e^2}{h}$ & $G_{xx}=n$ &  $G_{yy}(W-)$ & $G_{zz}(W-)$   \\
\cmidrule{1-4} 
%\multirow{2}{*}{\shortstack{6-terminal \\ resistance $R_{i j }/\frac{h}{e^2}$}} 
\multirow{2}{*}{6-terminal resistance $R_{ji}/\frac{h}{e^2}$} 
& $R_{xx}=0$ & $R_{yy}(B\mp) \ll 1$   & $R_{zz}(W+)\neq0$ \\ 
%\cline{2-4} 
\cmidrule{2-4} 
& $\tilde{R}_{xx}=R_{zx}=R_{yx}=\frac{1}{n}$ & $\tilde{R}_{yy} \approx G_{yy}^{-1}$  & $\tilde{R}_{zz} \neq R_{xz}\neq\frac{1}{n}$ 
\end{tabular}
\caption{Summary of the magnetotransport along three orthogonal directions. Current directions $i=x,y,z$ in the WSM with WPs along $k_z$-axis. Number of * indicates the stability against disorder: $i=y$ exhibits a moderate stability weaker than the QHE when $i=x$ but stronger than the vulnerable $i=z$ case. 
Major transport contributing channels are listed.
$n$ for quantized integer in the QHE. $G_{yy,zz}(W=0)$ is quantized to the number of contributing channels not related to QHE. Besides Hall and longitudinal resistances $R_{ji}$,
one has another resistance $\tilde{R}_{ii}$ between two current leads in 6-terminal measurement.
$W+$ ($W-$) means increasing (decreasing) as the disorder strength $W$ increases.
Small $R_{yy}$ originates from the separation both in $\bk$- and $\br$-space of the counter-propagating conducting channels.
$B\mp$ signifies the reversal from negative to positive magnetoresistance.}\label{Table:feature_summary}
\end{table*}

Since QHE is essentially a 2D phenomenon, it is an intriguing question whether the nature of this new QHE remains the same as in 2D\cite{Klitzing_1980,QHE1990,von_Klitzing_2020}. Surprisingly, dissimilar to what one would expect for a QHE-like phenomenon, there hides essential and intrinsic anisotropy in the magnetotransport, which can in general affect and even diminish the QHE. The clear QHE features when current flows transversely to the momentum-space WP alignment will be strongly suppressed as one deviates from this probing geometry. In the orthogonal setting with current parallel to WP alignment, the system does not show QHE and is the most susceptible to disorder. The origin can be traced back to an unexpected absence of bulk-boundary correspondence (BBC) and all the three directions play different roles: arc surface states from quantum anomalous Hall effect (QAHE) edge states are defined on momentum slices only between the WPs, and the role of green side surfaces is distinct from the front and back surfaces in Fig.~\ref{Fig:HallBar}(a). 
Besides revealing the fundamental difference between conventional QHE and this 3D WSM QHE, transport along the applied magnetic field shows a remarkable reversal of the well-known negative magnetoresistance due to chiral anomaly, which is attributed to a nontrivial competition between surface and bulk contributions.
To summarize, there are three directions, i.e., those of the applied magnetic field, the alignment of WPs, and that perpendicular to the previous two; the system exhibits an overlooked but essential anisotropy with respect to the transport direction relative to these three (Table.~\ref{Table:feature_summary}).

\begin{figure*}[hbt]
\includegraphics[width=17.cm]{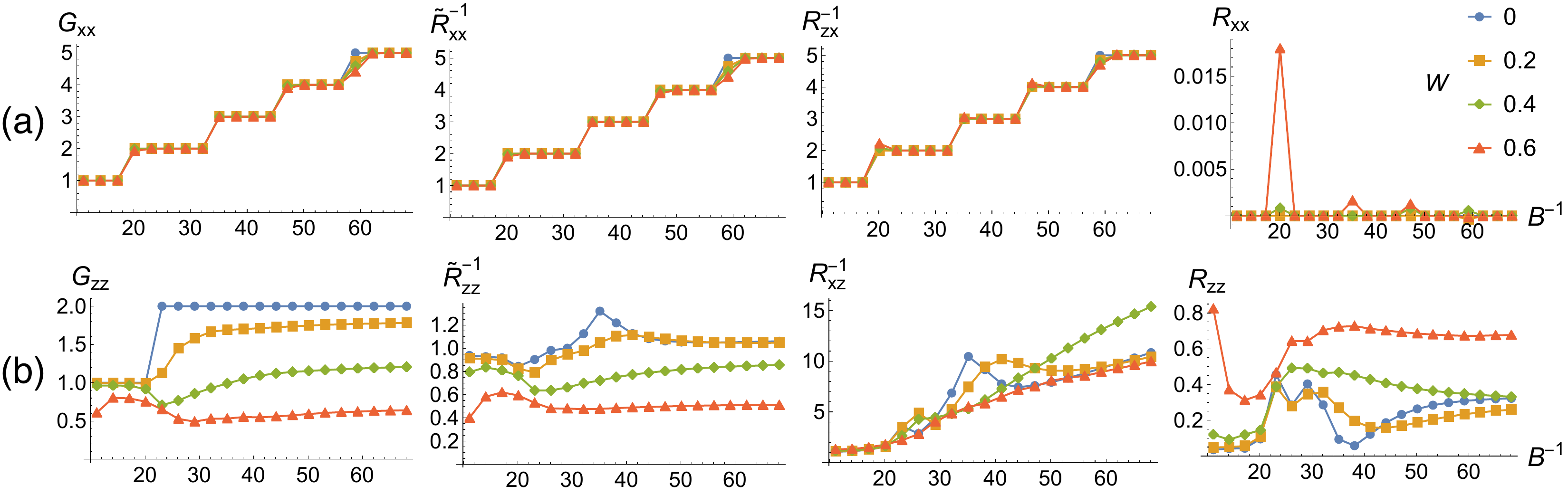}
\caption{Hall and longitudinal transport responses, including the 2-terminal conductance $G_{ii}$ and 6-terminal resistances for two orthogonal probing geometries, i.e., current flowing along (a) $i=x$ and (b) $i=z$. $R_{ii},R_{ji}$ are the longitudinal and Hall resistances; $\tilde{R}_{ii}$ is the resistance between current leads. Colors denote the disorder strength $W$. System size $L_x=L_z=37,L_y=28$. Parameters throughout this paper are $A=1,M=0.15,D_x=0.06,D_y=0.03,D_z=0.09,k_w=\pi/2$ and we set $e^2/h=1$.}\label{Fig:zxxz}
\end{figure*}

\section*{Results}

We consider the following minimal two-band model with two WPs $(0,0,\pm k_w)$ located on the $k_z$-axis and put the Fermi energy $\varepsilon_F=2D_z(1-\cos{k_w})$ at the WPs 
\begin{align}\label{eq:H}
\begin{split}
    H(\bk)=& \sum_{i=x,y,z} 2D_i (1-\cos{k_i})\sigma_0 + A(\sin k_x\sigma_x+\sin k_y\sigma_y) \\
    &+ 2M [(1-\cos{k_w})-\sum_{i=x,y,z}(1-\cos{k_i})]\sigma_z,
\end{split}
\end{align}
where we include mutually unequal $D_{x,y,z}$ for realistic anisotropy giving rise to curved Fermi arcs and nonvanishing response in every direction. Illustrated in Fig.~\ref{Fig:HallBar}(b), we use 2-terminal and 6-terminal Hall-bar probing geometries, where WSM current leads cover two surfaces and metallic voltage leads have limited size\cite{Datta1995,Datta2005}. Unless otherwise stated, we apply a magnetic field $\bB=B\hat{y}$ via the Peierls substitution $\bk\rightarrow\bk+e\bA$ %, where the gauge potential $\bA$ is chosen such that it is independent of the position along current direction. 
and average over an onsite disorder uniformly distributed in $[-W/2,W/2]$. For a current flowing along $i$-axis, apart from the 2-terminal conductance $G_{ii}$, the 6-terminal Hall bar measures either the Hall ($j\neq i$) or longitudinal ($j=i$) resistance $R_{ji}$. Another $\tilde{R}_{ii}$ between leads 5,6 would give the identical Hall resistance $R_{ji}$ if the BBC in conventional QHE was preserved, i.e., contributed by the quantized contact resistance from edge states as $G_{ii}$. To calculate the transmission probability in the Landauer-B\"uttiker formula, we adopt the wavefunction scattering matrix approach equivalent to the nonequilibrium Green's function\cite{Fisher1981,Jiang2009,Groth2014}.

\begin{figure*}[hbt]
\includegraphics[width=16.cm]{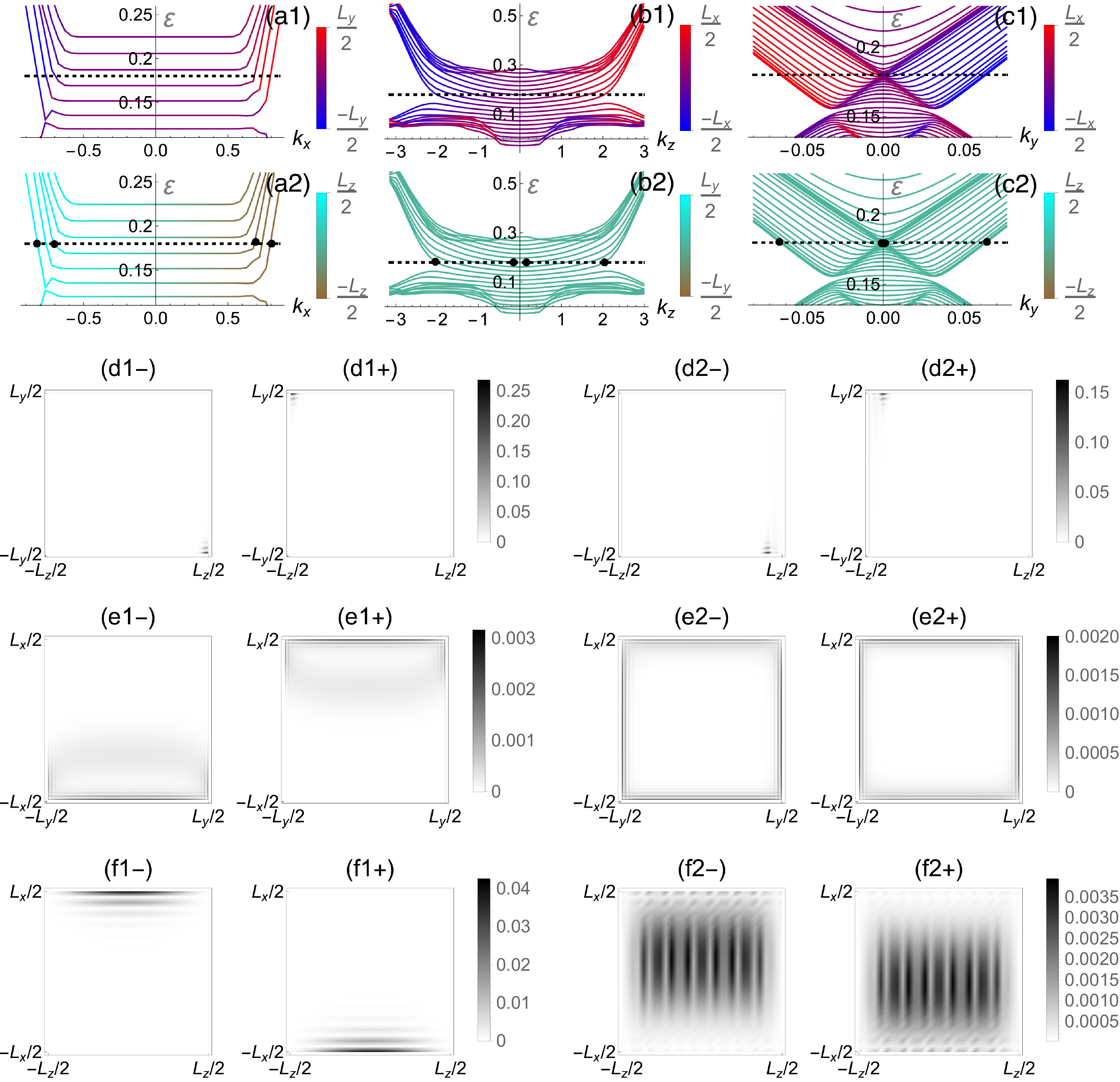}
\caption{(a-c) Low-energy bands under magnetic field with (a) periodic $x$-axis and $L_y=L_z=60,B=0.025$ (b) periodic $z$-axis and $L_x=L_y=90,B=0.025$ (c) periodic $y$-axis and $L_x=L_z=30,B=0.1$. The main features are stable against large enough system sizes. Fermi energy $\varepsilon_F=0.18$ is indicated by the dashed line. Color scales indicate the real-space position expectation value $\braket{y},\braket{z},\braket{x},\braket{y},\braket{x},\braket{z}$ respectively for (a1,a2,b1,b2,c1,c2) of the corresponding wavefunction. Note that $\braket{y},\braket{z}$ are mostly close to zero in (b2,c2).
(d-f) Wavefunction probability distribution in three cross-sections (d) $yz$ (e) $yx$ (f) $zx$ respectively corresponding to (a,b,c). For (d), wavefunctions from two representative bands crossing the Fermi level in (a), typically with (d1) largest or (d2) smallest $k_F$ along $k_x$ as indicated by black dots in (a2), are shown; within (d1), for instance, (d1$\pm$) respectively shows the state at positive/negative $k_x=\pm k_F$ of that particular band. (e,f), labelled and sorted in the similar manner, are from the black dots in (b2,c2).}\label{Fig:bands}
\end{figure*}

\subsection*{Suppression of QHE in an orthogonal geometry}

In Fig.~\ref{Fig:zxxz}, we show the conductance and resistance for two orthogonal probing geometries. While the $zx$-geometry (current along $x$ and voltage leads extend along $z$) exhibits clear and consistent QHE features robust against disorder\cite{Wang2017a,Li2020a}, the $xz$-geometry sees entirely distinct behavior. i) The quantized plateaus disappear in any of $G_{zz},\tilde{R}_{zz},R_{xz}$ except $G_{zz}$ without disorder. ii) The vanishing longitudinal resistance $R_{zz}$ becomes finite and overall increases with disorder strength in a conventional manner. iii) The stability against disorder in each signal largely reduces to a minimal level. This indicates that the conducting channels contributing to the transport along two orthogonal directions bear some fundamental difference. 
Note that this does not contradict the Onsager reciprocity discussed in Supplemental Information (SI Sec.~I).

In fact, without disorder one can visualize the conducting channels thanks to the conserved momentum along the respective current direction. Fig.~\ref{Fig:bands} shows the relevant band structure and the corresponding representative cross-sectional wavefunction probability distribution. In Fig.~\ref{Fig:bands}(a), the projection of both WPs coincide at $k_x=0$. The formation of LLs in a range across $\varepsilon_F$ signifies their origin from the surface Fermi arcs, since the bulk density of states vanishes at $\varepsilon_F$. Then the upward-bending edge of these LLs exactly serves as the red diagonal QHE edge states in Fig.~\ref{Fig:HallBar}(a) under protection from backscattering, which is also readily seen in Fig.~\ref{Fig:bands}(d). In Fig.~\ref{Fig:bands}(b), two WPs at $k_z=\pm\pi/2$ remain separated and are connected by the arc states discretized by finite $L_{x,y}$. 
Fig.~\ref{Fig:bands}(e1) with large $k_F$ exhibits counter-propagating D-shaped wavefunctions connecting the top and bottom via the side surface and the bulk, whose semiclassical trajectory has previously been noticed\cite{Li2020a}. 
In Fig.~\ref{Fig:bands}(e2) with small $k_F$, the top and bottom surfaces and two side surfaces are almost fully connected. In fact, as one plots the conducting wavefunctions with decreasing $k_F$, Fig.~\ref{Fig:bands}(e1) gradually deforms into Fig.~\ref{Fig:bands}(e2).
Most importantly, there is no LL formation along $k_z$, which dictates the disappearance of QHE together with related features. 
Large system sizes can at most give few fragile quasiplateaus well below the accuracy and robustness of QHE (SI Sec.~II). It partially results from the conventional mesoscopic channel quantization as $G_{zz}(W=0)$ and may grow in large samples due to denser discretization. 
The results are summarized in Table.~\ref{Table:feature_summary}. See also SI Sec.~I for the complete resistance tensor. Transport along the magnetic field direction $\hat{y}$ in relation to Figs.~\ref{Fig:bands}(c,f) will be described later.

Surface theory can help gain further insights\cite{Okugawa2014,Zhang2016a,Wang2017a,Borchmann2017}. For the top and bottom $y=\pm L_y/2$ surfaces around $k_x=0$, we obtain the effective Hamiltonian (SI Sec.~III)
\begin{equation}\label{eq:h_surface}
    h_\pm(k_x,k_z)=\varepsilon_0 \pm v'k_x+D_x'k_x^2- 2D_z'\cos{k_z}
\end{equation}
for $k_z$ between WPs, where $\varepsilon_0=2D_z-2D_y\cos{k_w},v'=A\sqrt{M^2-D_y^2}/M$ and $D_x'=D_x-D_y,D_z'=D_z-D_y$. The top and bottom Fermi arcs are respectively in the $k_x>0$ and $k_x<0$ regions due to positive $D_z'$. 
Considering transport along $x$-axis, gauge choice $\bA=Bz\hat{x}$ and conserved $k_x$ is able to form LLs from essentially the same quasi-1D $z$-axis cosine gas.
%These two top and bottom gases' LLs do not seem to be affected by the constraint or the unclosed arc. In fact, the constraint picture enters here only in that the top surface LLs account for the right half of the flat bands (kx>0) and the bottom for the left part. Therefore, the full flat band requires both top and bottom and hence also the bulk chiral LLs for bridging.
Semiclassically, LL states originate from electrons traversing the Weyl orbit previously described with Fig.~\ref{Fig:HallBar}(a), which consists of arc trajectories connected by bulk chiral LLs. 
The metallic $\pm L_x/2$ green side surfaces in Fig.~\ref{Fig:HallBar}(a), which host QAHE circulating states 
and disperses along $y,z$-directions, are merged in the current leads. In fact, around the $\pm L_y/2$ surface, arc LL bands bent upward in Fig.~\ref{Fig:bands}(a) near the $\mp L_z/2$ edge will just become the $\pm\hat{x}$-propagating red diagonal edge channel in Fig.~\ref{Fig:HallBar}(a), because of the correspondence between $k_x$ and the guiding center along $z$-axis. 
Figs.~\ref{Fig:bands}(d1,d2) clearly show such diagonal edge state in the $yz$-plane.
This means that the $h_\pm$ contributions to edge LL formation are largely separated in both $k_x$ and real $z$ and do not directly interfere.

However, for transport along $z$-axis with gauge choice $\bA=-Bx\hat{z}$ and conserved $k_z$, the top and bottom quasi-1D $x$-axis gases are not 'decoupled' in the previous simple manner. Required by the measurement geometry, the nontrivial green side surfaces in Fig.~\ref{Fig:HallBar}(a) are always present and host similar states as Eq.~\eqref{eq:h_surface}; their highly nonlocal QAHE circulating states naturally extend into the top and bottom surfaces and play in part the role of connecting states from $h_\pm$ under magnetic field.
Therefore, a magnetic arc state at $\varepsilon_F$ lives on \textit{both} top and bottom surfaces and exhibits $\braket{y}\sim 0$ in Fig.~\ref{Fig:bands}(b2), and possesses a much weaker $\braket{x}$-asymmetry in Fig.~\ref{Fig:bands}(b1) than Fig.~\ref{Fig:bands}(a2). These are readily visualized within the $yx$-plane in Fig.~\ref{Fig:bands}(e) as previously noted.
Indicated by these results, the circulating or top-bottom inseparable nature forces hybridization with the bulk and especially metallic side surfaces and generates dispersiveness in Fig.~\ref{Fig:bands}(b); there is thus no flat bulk gap essential to QHE but hybridized states that allow for scattering. Note that the metallicity of the side surface, e.g., its $z$-axis dispersion, is indispensable to the 3D QHE as it generates the curved Fermi arcs.
Fundamentally, such contrasting behavior between two geometries is the consequence of that the WSM topology chooses a direction: WPs are aligned along $k_z$-axis. Phenomena possibly related to this have been noticed in the second-harmonic generation and arc optical conductivity\cite{Wu2016,Shi2017}.

\begin{figure}[hbt]
\includegraphics[width=8.6cm]{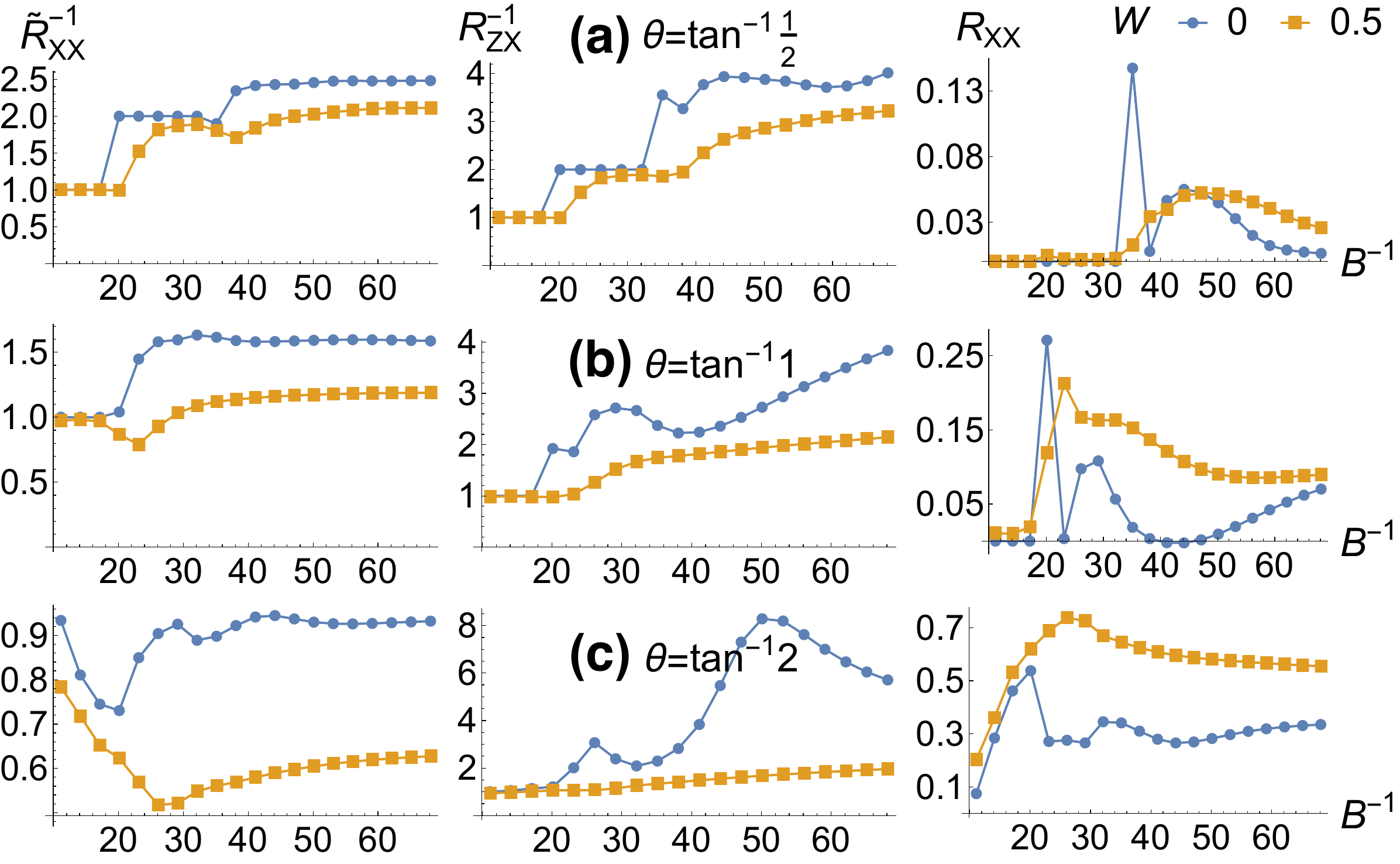}
\caption{Hall and longitudinal resistances $\tilde{R}_{ii},R_{ji},R_{ii}$ for 6-terminal measurement. The current direction $i=X$ (together with the orthogonal $j=Z$ direction) of the Hall bar is rotated with respect to $\hat{y}$ from the original $x$-axis by a commensurate angle $\theta$. 
System size $L_y=24$ and $L_{X,Z}$ use the corresponding commensurate length closest to $38$ in three rotation angles (a,b,c) in ascending order. Colors denote the disorder strength $W$.}\label{Fig:rotate_main}
\end{figure}

\subsection*{Rotating the system with respect to \texorpdfstring{$y$}{y}-axis}
In realistic WSM experiments, alignment of WPs as generic band crossings could easily be oblique to the natural crystallographic growth geometry and also the Hall bar, e.g., due to hoppings beyond nearest neighbour. To address this, we consider a rotation with respect to $\hat{y}$ of the Hall-bar system by a commensurate angle $\theta=\arctan(m_x,m_z)$: the new supercell includes $m_i$ cells along $i$-axis for $i=x,z$ [see SI Fig.~S4]. Accordingly, we rotate the gauge $\bA(\theta)$ together (SI Sec.~IV), denote the rotated $x,z$-axes as $X,Z$ and keep the current flowing along $X$. When $\theta=0,\pi/2$, it reproduces the foregoing $zx$- or $xz$-geometry. As shown in Fig.~\ref{Fig:rotate_main} and additional cases in SI Sec.~II, the transport responses exhibit a significant anisotropy. i) Higher QHE plateaus start to disappear as $\theta$ increases from 0. ii) Beyond $\pi/4$ basically no plateaus can survive. iii) The remaining plateaus and longitudinal resistances are much more susceptible to the same level of disorder than the $zx$-geometry in Fig.~\ref{Fig:zxxz}(a). Even in Fig.~\ref{Fig:rotate_main}(b) the first plateau with disorder actually suffers a huge 2\% deviation from perfect quantization, compared with $10^{-13}$-accuracy in Fig.~\ref{Fig:zxxz}(a). %Larger system sizes do not alleviate the above features.
Note also that in a conventional QHE, e.g., a square-lattice electron gas, none of these features are present and the plateaus are always robust against such rotation although the system itself only possesses a $C_4$ rotation symmetry (SI Sec.~II).

This anisotropy or strong $\theta$-dependence displays a crossover behavior in between the two previous cases. With an oblique $\theta>0$, transport along $X$ is no longer contributed solely by the well-defined arc LL edge states; instead, there always exists a finite projection in the $z$-direction transport, which is based on the foregoing surface hybridized states without LL formation. As $\theta$ increases, the $z$-axis backscattering affects more and more the $X$-axis current, which does not play a role when $\theta=0$. Therefore, the QHE will deteriorate to some extent with any finite $\theta$ because the formation and robustness of plateaus and the vanishing longitudinal resistance indispensably rely on an exclusive LL edge-state transport.

\subsection*{Absence of conventional bulk-boundary correspondence}

An important aspect of the conventional QHE consists in the BBC, where the number of conducting edge channels can be uniquely determined by the bulk topological Chern number\cite{ChernTKNN1,ChernTKNN2}. Combining the foregoing physical picture and concrete transport responses, we conclude that the QHE in 3D WSM does not possess an usual BBC. Firstly, a direct consequence of BBC in conventional QHE is the aforementioned identity $\tilde{R}_{ii}=R_{ji}$ when directions $i,j$ with $i\neq j$ are perpendicular to magnetic field, which is violated as long as $i\neq x$ or $\theta\neq0$ as shown in Figs.~\ref{Fig:zxxz} and \ref{Fig:rotate_main}. Secondly, the QHE hereof largely relies on the presence of surface arc states under magnetic field, where $\varepsilon_F$ pinpointed at the WP crosses arc LLs only in Fig.~\ref{Fig:bands}(a) and bulk states participate in a supporting way. This renders it physically infeasible to define a \textit{bulk} topological invariant. %The possible link to higher-order topology was otherwise noted\cite{Chen_2021}. 
Importantly, the present phenomenon is distinct from the earlier 3D QHE scheme characterized by Chern numbers defined for three orthogonal planes, which is purely a bulk magnetic band property caused by the charge-density-wave gap etc.\cite{Avron1983,Halperin1987,Montambaux1990,Tang2019,Qin2020,Galeski_2021}.  Thirdly, to reassure this viewpoint, one can actually calculate the conductivity $\sigma_{zx}$ from the lattice model via the Kubo-Bustin formalism with kernel polynomial method, where operator evaluation is restricted deep \textit{inside} the bulk\cite{Weise2006,Garcia2015,Fan2021}. While it fails to display the present QHE, the same calculation well shows both the conventional QHE in a 2D electron gas and the expected AHE $\sigma_{xy}$ proportional to $|2k_w|$ in this WSM system (SI Sec.~V). This is simply because the latter two cases bear the topological BBC.

\begin{figure}[hbt]
\includegraphics[width=8.6cm]{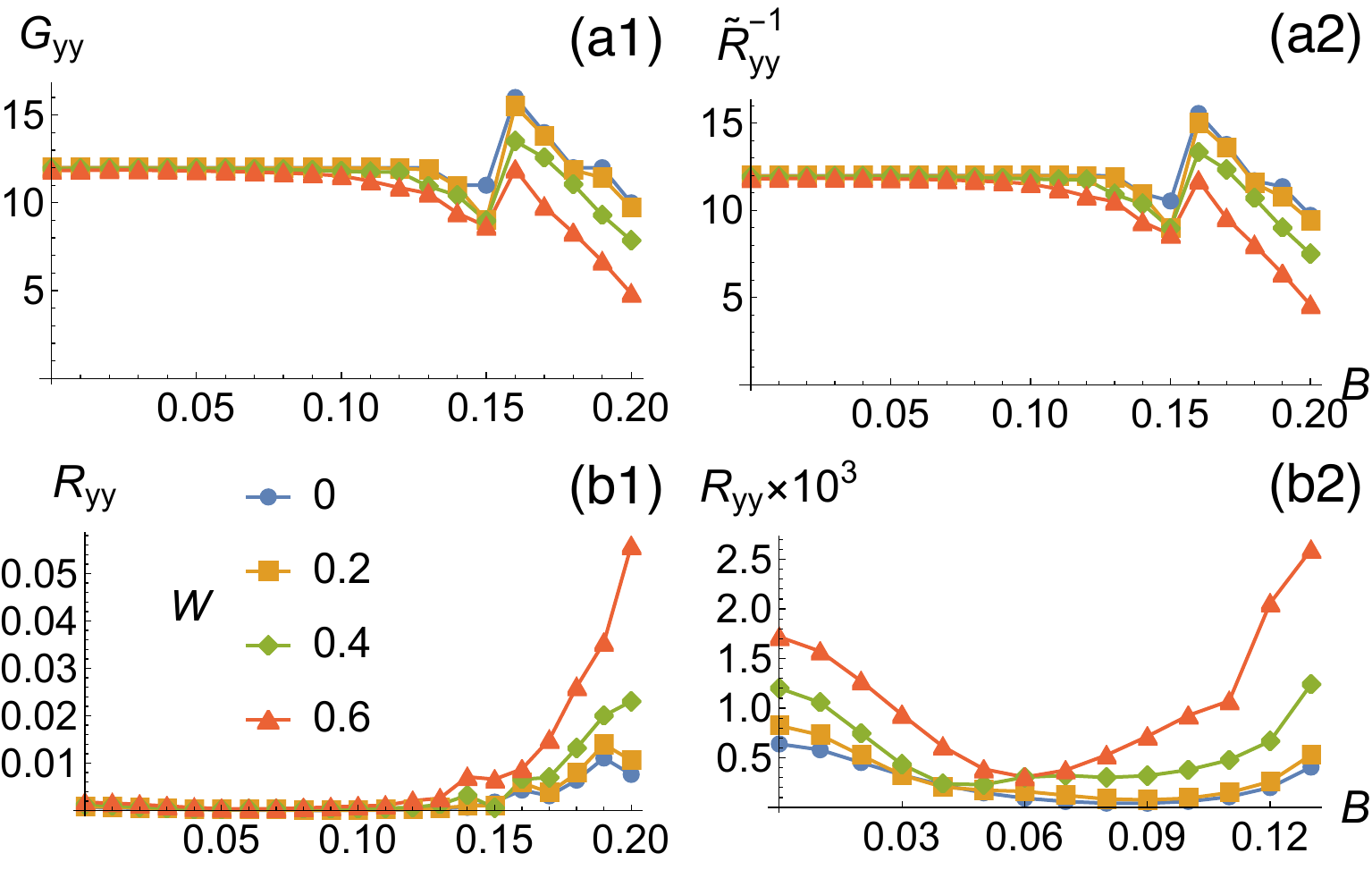}
\caption{Magnetotransport parallel to magnetic field along $y$-axis. (a1,a2) Conductance $G_{yy}$ and resistance  $\tilde{R}_{yy}$. (b2) zooms in the low-field region of the longitudinal resistance $R_{yy}$ in (b1) where the reversal from negative to positive magnetoresistance happens. System size $L_x=L_z=30,L_y=100$.}\label{Fig:ysignal}
\end{figure}

\subsection*{Transport parallel to the magnetic field}
The remaining $y$-axis magnetotransport also has notable features. This configuration is relevant to the negative magnetoresistance due to chiral anomaly in WSM, where chiral pumping via bulk chiral LLs between WPs is at work\cite{Nielson-Ninomiya2,CME1,Burkov2014}.
As shown in Fig.~\ref{Fig:ysignal}, $G_{yy},\tilde{R}_{yy}$ consistently remain nearly constant until a large field destroys the contributing QAHE accumulated between WPs. Before that, the robustness against disorder is intermediate in between the previous two orthogonal $zx$- and $xz$-geometries; relatedly, the longitudinal $R_{yy}\ll1$ but is not vanishing and $\tilde{R}_{yy}$ approximately follows $G_{yy}^{-1}$ as indicated in Table.~\ref{Table:feature_summary}.
This intermediate stability results from a combination of $\bk$- and $\br$-space separation unique to this transport direction.
The relevant conducting channels along $y$-axis include two parts in Fig.~\ref{Fig:bands}(c1). i) Red and blue QAHE edge states near the $x=\pm L_x/2$ side surfaces are spatially separated. ii) Two sets of counter-propagating purple bulk chiral LLs that cross at $k_y=0$ where the projection of two WPs coincides; they are distributed inside the system bulk and also near the side surface. Figs.~\ref{Fig:bands}(f1,f2) respectively show the probability distribution of these two types of states in the $zx$-plane.

Intriguingly in Fig.~\ref{Fig:ysignal}(b1) and the magnified (b2), the negative magnetoresistance is very weak and holds only up to a medium field strength as disorder can strongly suppress and reverse it to positive magnetoresistance. The reversal point moves to smaller field strength with either larger $W$ or $L_y$. Note that this is entirely absent in the other geometry when $\bB$ and current $I$ both are along $z$-axis, i.e., parallel to the WP alignment, where negative magnetoresistance persists all along (SI Sec.~I). This robust negative magnetoresistance means that, although the $\bk$-space separated bulk chiral LLs are subject to large-momentum-transfer disorder scattering, the increasing number of these chiral channels with larger $B$ guarantees it. 
Physically, the main difference in Fig.~\ref{Fig:ysignal}(b) from the $\bB\| I\|\hat{z}$ situation is the contribution from the aforementioned nontrivial green side surfaces in Fig.~\ref{Fig:HallBar}(a) that host topological states conducting along $y$-axis. These dispersive surface states can themselves be backscattered and especially to the counter-propagating chiral LLs also present at the side surfaces, which is made easier and more frequent by the denser guiding center lattice at larger $B$: the side-surface contribution is purely positive magnetoresistance.
Remarkably in this competition, eventually the surface backscatterings prevails and dominantly contributes to the longitudinal resistance $R_{yy}$. In a 3D bulk sample with $L_y$ comparable to or even larger than the mean free path, such a positive magnetoresistance will naturally be expected to occur and it reveals the crucial but overlooked surface magnetotransport contribution in WSM.

\section*{Discussion}
%\mytitle{Summary}.--
%We examine the 3D QHE and magnetotransport proposed in WSM systems from the viewpoint of Hall-bar measurement that is most experimentally relevant. We reveal an essential but previously unaddressed fully 3D anisotropy due to the intrinsic WP alignment and the absence of BBC. While the transport along magnetic field shows a reversal of the well-known negative magnetoresistance, the QHE becomes susceptible to the probing geometry in the plane perpendicular to magnetic field, which disappears in a crossover manner when the current direction varies from transverse to parallel to the WP alignment. These findings will deepen the understanding of the peculiar surface-bulk-hybrid nature of magnetotransport in WSM and are also expected to provide immediate and key intuition to the experimental investigation.
%Our findings establish the WSM QHE system as a novel quantum state of matter of highly nontrivial 3D fully anisotropic surface-bulk-hybrid magnetotransport and bear no conventional counterpart in the ordinary 2D QHE. 
We examine the 3D QHE and magnetotransport in WSM systems from the experimentally most relevant viewpoint of Hall-bar measurement.
Our findings establish the system as a novel quantum state of matter of highly nontrivial fully 3D anisotropy and reveal the peculiar surface-bulk-hybrid nature of magnetotransport with no conventional counterpart, e.g., in the ordinary 2D QHE.
It is expected to provide immediate and key insights to the experimental investigation.
Although we mainly concern the case with one single pair of WPs as the physically most clarifying situation, the general consideration can be readily extended to more complicated cases with multiple pairs of WPs, regardless of magnetic or nonmagnetic WSM (SI Sec.~VI). %The resultant response features will depend on the Fermi arc connectivity of those WPs, which is determined by the bulk topology and also surface conditions\cite{Kim_2016,Morali_2019,Lv2021}. Let's exemplify the situation with one more pair of WPs connected by Fermi arc in addition to the original pair we discussed. When this second pair of WPs projected onto the orange top or bottom surface is largely parallel to that of the first pair, they contribute additively to the red arc LL edge states in Fig.~\ref{Fig:HallBar}(a); the 3D anisotropy in QHE and other magnetotransport is naturally expected to remain the same as discussed. On the other hand, when the projection of the second pair is considerably oblique to the first pair, e.g., the white front and back surfaces will become metallic and nontrivial in a similar manner to the green side surfaces in Fig.~\ref{Fig:HallBar}(a), it can diminish the red edge states and is related to the rotated situation we discussed. Therefore, transport in $xz$-plane will in general have weakened or even destroyed QHE while the longitudinal magnetoresistance along $B\hat{y}$ is possible to have more enhanced reversal effect due to the extra nontrivial surfaces. 

Among experimental reports of quantized transport in the closely related Dirac semimetal Cd$_3$As$_2$, a few provide supportive evidence of the role of Weyl orbit via thickness modulation and dual-gate modulation between top and bottom surfaces\cite{Zhang2018,Nishihaya2021,Zhang2021a}. In the experiment with a nanobelt grown by chemical vapor deposition, current along the in-plane $[1\bar{1}0]$ direction flows through the sample with surfaces normal to the magnetic field in the $[112]$ direction\cite{Zhang2018}. One can identify them respectively as $\hat{x}$ and $\hat{y}$ directions in our discussion. The Dirac points along $[001]$ direction (oblique but inside the $yz$-plane) lead to pairs of Fermi arcs along $z$-axis on the top and bottom surfaces in a similar manner as in Fig.~\ref{Fig:HallBar}(a). The front and back surfaces can also host surface states propagating along $x$-axis but do not affect the red diagonal edge states in Fig.~\ref{Fig:HallBar}(a). Hence it is akin to the $zx$-geometry with QHE as discussed. 
In the other thin-film experiment using solid-phase epitaxy, the in-plane current direction is presumed to include the $[1\bar{1}0]$ and $[11\bar{1}]$ directions due to the presence of two possible domains\cite{Nishihaya2021}. Therefore, this situation corresponds to the rotated case effectively and is still possible to exhibit lower quasiplateaus in the mesoscopic transport as we discussed. Indeed, the quantization bears a much more extended and robust appearance and the longitudinal resistivity is even closer to zero in the former nanobelt experiment\cite{Zhang2018}, which is consistent with our study.
Further experiments in especially WSM under various geometries we proposed will certainly be helpful to firmly observing the intriguing 3D anisotropic effects.

%In summary, we examine the 3D QHE and magnetotransport in WSM systems from the viewpoint of Hall-bar measurement that is most experimentally relevant. We find an essential but previously unaddressed fully 3D anisotropy due to the intrinsic WP alignment and an external magnetic field perpendicular to that alignment. The QHE becomes susceptible to the probing geometry in the plane perpendicular to magnetic field, which disappears in a crossover manner when the current direction varies from transverse to parallel to the WP alignment. This is traced back to an unexpected absence of conventional BBC. Not only revealing the fundamental difference between conventional 2D QHE and the 3D WSM QHE, we also find that the transport along the applied magnetic field shows an intriguing reversal of the well-known negative magnetoresistance due to chiral anomaly, which is attributed to a nontrivial competition between surface and bulk contributions. These findings establish the 3D WSM QHE system as a novel quantum state of matter with no conventional counterpart, revealing the peculiar surface-bulk-hybrid nature of magnetotransport herein, and are particularly expected to provide immediate and key insights to the experimental investigation.

\section*{Associated content} 
The Supporting Information is available free of charge at
https://pubs.acs.org/doi/10.1021/acs.nanolett.

Other geometries and additional data, details of model analysis, conductivity calculation, and discussion of multiple pairs of Weyl points.

\begin{acknowledgments}
X.-X.Z. is thankful for the helpful discussion
with H. Li,
M. Uchida, P. Perez-Piskunow, C. Zhang, and Y. Tokura and also for the computation facility coordinated by H. Isobe. This work was supported by JSPS KAKENHI (No.~18H03676) and JST CREST (Nos. JPMJCR1874 \& JPMJCR16F1). X.-X.Z was partially supported by RIKEN Special Postdoctoral Researcher Program.
\end{acknowledgments}\mycomment{\Yinyang}

%\nocite{*}

\bibliography{reference.bib}  % The references (bibliography) information are stored in the file named "Bibliography.bib"

\newpage
\onecolumngrid
\begin{figure}[hbt]
\includegraphics[width=8.7cm]{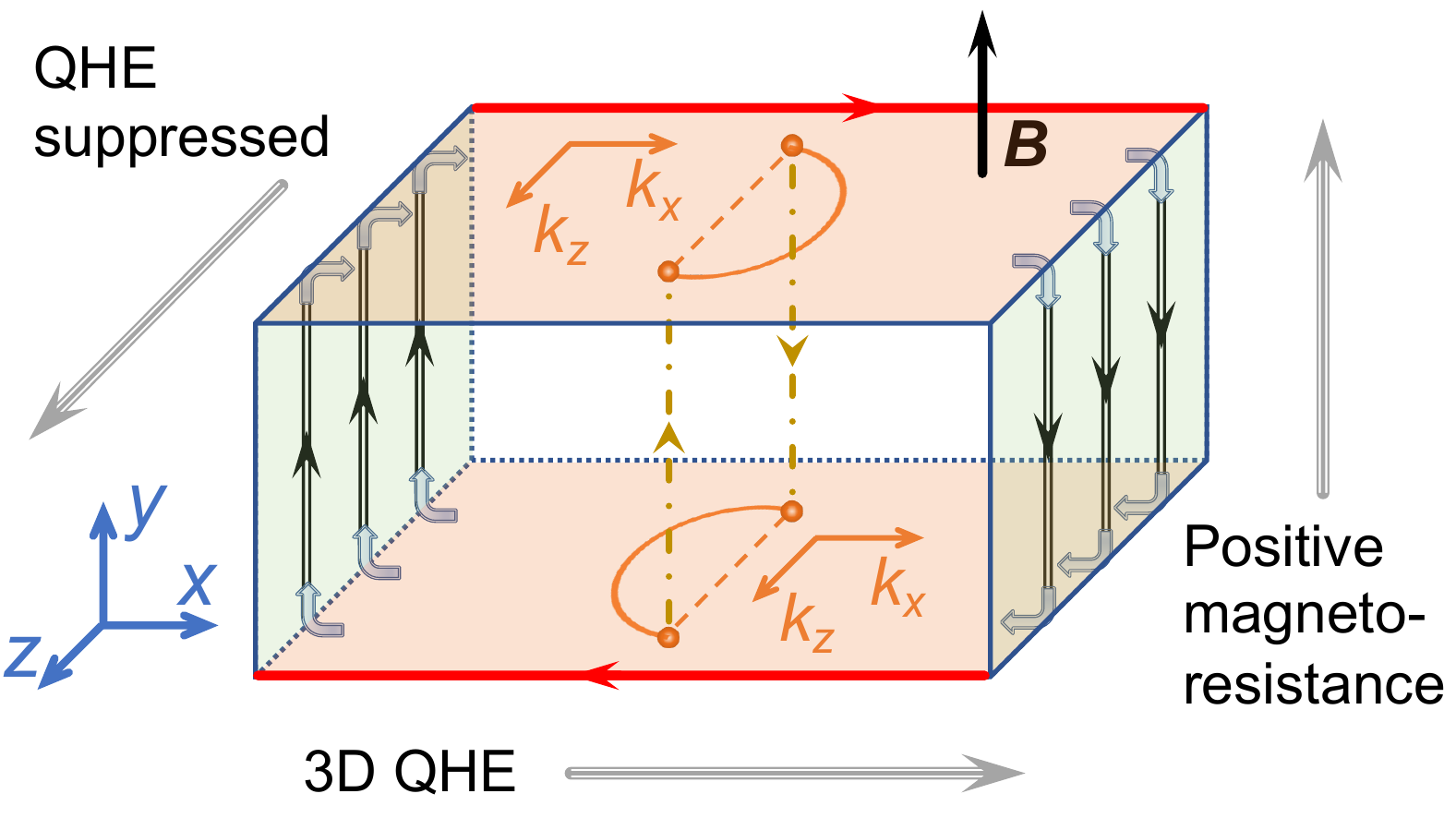}
\caption{For Table of Contents Only}\label{Fig:TOC}
\end{figure}

\end{document}